\documentclass{PoS}
\usepackage{amsfonts}
\usepackage{mathrsfs}
\usepackage[ansinew]{inputenc}      
\usepackage{amssymb}                
\usepackage{amsmath}                
\usepackage[]{graphicx}
\usepackage{mathrsfs}
\usepackage{verbatim}
\usepackage{overpic}
\usepackage{xspace}
\usepackage{lineno}
\usepackage{enumitem}
\usepackage[font=small,skip=3mm]{caption}
\usepackage{setspace}
\usepackage{array}
\usepackage{lscape}
\usepackage{rotating}

\newcommand{\ttbar}{\ensuremath{t\bar{t}}\xspace}
\newcommand{\etmiss}{\ensuremath{E \kern-0.6em\slash_{\rm T}}\xspace}
\newcommand{\etmissx}{\ensuremath{E \kern-0.6em\slash_{\rm x}}\xspace}
\newcommand{\etmissy}{\ensuremath{E \kern-0.6em\slash_{\rm y}}\xspace}

\newcommand{\ljets}{\ensuremath{\ell\!+\!{\rm jets}}\xspace}

\newcommand{\etal}{\textit{et~al.}}

\newcommand{\GeV}{\ensuremath{\textnormal{GeV}}\xspace}
\newcommand{\TeV}{\ensuremath{\textnormal{TeV}}\xspace}

\newcommand{\dif}{\ensuremath{{\rm d}}}

\newcommand{\fb}{\ensuremath{{\rm fb}^{-1}}\xspace}
\newcommand{\mw}{\ensuremath{M_W}\xspace}
\newcommand{\mt}{\ensuremath{m_t}\xspace}

\newcommand{\kjes}{\ensuremath{k_{\rm JES}}\xspace}

\newcommand{\pt}{\ensuremath{p_T}\xspace}

\newcommand{\stt}{\ensuremath{\sigma_{t\bar t}}\xspace}
\newcommand{\afb}{\ensuremath{A_{\rm FB}}\xspace}
\newcommand{\afbtt}{\ensuremath{A_{\rm FB}^{\ttbar}}\xspace}
\newcommand{\afbl}{\ensuremath{A_{\rm FB}^{\ell}}\xspace}

\title{Recent results in the top quark sector\\
from the D0 experiment}

\ShortTitle{Top results from D0}

\author{\speaker{Oleg Brandt}\thanks{FERMILAB-CONF-14-564-E}\\
        Kirchhoff-Institut f\"ur Physik, Im Neuenheimer Feld 227,\\
        69120 Heidelberg, Germany\\
        E-mail: \email{oleg.brandt@kip.uni-heidelberg.de}}

\abstract{
In these proceedings, I review recent measurements in the top quark sector in $p\bar p$ collisions at a centre-of-mass energy of $\sqrt s=1.96$ TeV in Run II of the Fermilab Tevatron Collider using the D0 detector. I will present the differential measurement of the $t\bar t$ production cross section and the Tevatron combination of inclusive $t\bar t$ cross section measurements; the first evidence of the production of single top quarks in the $s$-channel by D0 and the observation in combination with CDF. Furthermore, I will review the measurements of the forward-backward asymmetry in $t\bar t$ events, and conclude with the world's most precise single measurement of the top quark mass, which is a fundamental parameter of the standard model, and present the Tevatron combination, which is the world's most precise determination of the top quark mass, with a relative precision of 0.37\%.

}

\FullConference{XIIth International Conference on Heavy Quarks \& Leptons 2014,\\
		25-29 August 2014\\
		Schloss Waldthausen, Mainz, Germany}

\begin{document}

\section{Introduction}

Since its discovery~\cite{bib:topdiscovery}, the determination of the properties of the top quark has been one of the main goals of the Fermilab Tevatron Collider, recently joined by the CERN Large Hadron Collider. Observation of the electroweak production of single top quarks was presented only two years ago~\cite{bib:singletop}. The large top quark mass~\cite{bib:mt_tev} and the resulting Yukawa coupling of about $0.996\pm0.004$ in the standard model (SM) indicates that the top quark could play a crucial role in electroweak symmetry breaking. A precise measurement of the production mechanism of the top quark and of its intrinsic properties test the consistency of the SM and could hint at physics beyond the standard model (BSM). The top quark mass \mt is a fundamental parameter of the SM and has to be determined experimentally. Together with the mass of the $W$ boson \mw, and the mass of the Higgs boson it can be used as an internal consistency check of the SM. Furthermore, \mt dominantly affects the stability of the SM Higgs potential, which has related cosmological implications~\cite{bib:vstab}. 

I will present the differential measurement of the \ttbar production cross section and its Tevatron combination; the first evidence of the production of single top quarks in the $s$-channel by D0 and the evidence in combination with CDF, the measurements of the forward-backward asymmetry in \ttbar events, which can be experimentally uniquely accessed in the $CP$-invariant $p\bar p$ initial state at the Tevatron, and conclude with the world's most precise single measurement of \mt. All D0 results on the top quark can be found in~Ref.~\cite{bib:topresd0}, while the top quark measurements at the LHC are covered in Refs.~\cite{bib:lhc}. Most measurements presented here are based on $\approx10~\fb$ of $p\bar p$ collision data recorded by the D0 detector in Run II of the Fermilab Tevatron Collider.

\section{Top quark production at hadron colliders}

At the Tevatron, top quarks are dominantly produced in pairs through the strong interaction. The dominant production process is $q\bar q\to\ttbar$, which contributes about $85\%$ of the total \ttbar production cross section \stt, while at the LHC the $gg\to\ttbar$ reaction dominates with about $85\%$. Another striking feature of the $p\bar p$ initial state at the Tevatron is that it represents an eigenstate of the $CP$ transformation. The $s$-channel contributes $\approx30\%$ of the total production cross section of single top quarks $\sigma_t$ at the Tevatron, and is therefore experimentally easier to access than at the LHC, where it contributes only $\approx5\%$.
In the SM, the top quark decays to a $W$ boson and a $b$ quark nearly 100\% of the time. Thus, $\ttbar$ events decay to a $W^+W^-b\bar b$ final state, which is further classified according to the $W$ boson decays into ``dileptonic'', ``\ljets'', or ``all--jets'' channels. Accordingly, single top production in the $s$-channel proceeds dominantly through $q\bar q\to t\bar b$~\cite{bib:foot1}, while the $qg\to q't\bar b$ process dominates in the $t$-channel.

\section{$\boldsymbol{\ttbar}$ production cross section} 

\begin{figure}
\centering
\begin{overpic}[clip,height=4.5cm]{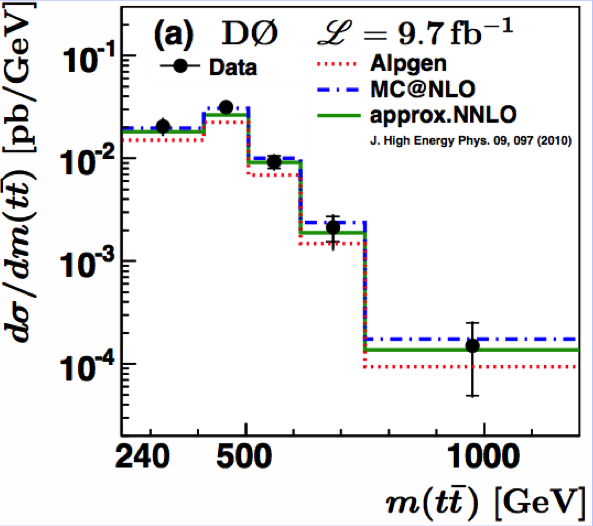}
\put(-1,2){(a)}
\end{overpic}
\begin{overpic}[clip,height=4.5cm]{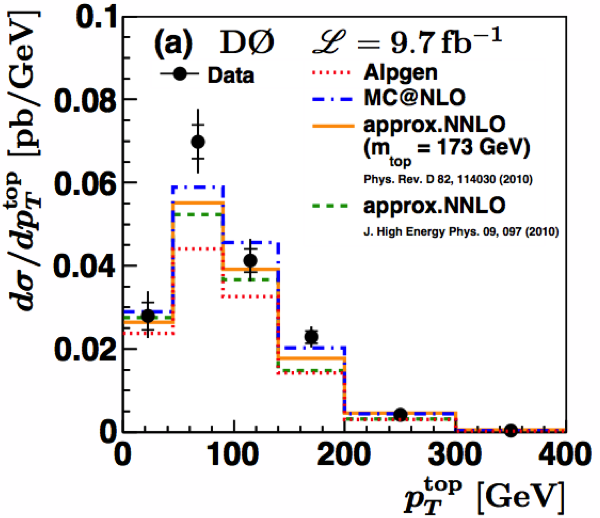}
\put(-1,2){(b)}
\end{overpic}
\caption{\label{fig:xsec_tt_diff}
{\bf(a)} The distribution of $\dif\stt/\dif m_{\ttbar}$ after background subtraction and regularised unfolding measured by D0 in the \ljets channel using 9.7~\fb of data~\cite{bib:xsec_tt_diff}, compared with theory predictions.
{\bf(b)} Same as (a), but for $\dif\stt/\dif \pt^{\rm top}$. Both $t$ and $\bar t$ contribute in each event in~(b).
}
\end{figure}

D0 recently measured $\stt$ differentially in various kinematic distributions in the \ljets chanel using 9.7~\fb of data~\cite{bib:xsec_tt_diff}. After subtraction of background contributions, the distributions are corrected for detector resolution and acceptance through a regularised matrix unfolding. Two representative results are shown in Fig.~\ref{fig:xsec_tt_diff}~(a) for $\dif\stt/\dif m_{\ttbar}$ and in~(b) for $\dif\stt/\dif \pt^{\rm top}$. The SM predictions considered are able to describe the differential dependence up to an overall normalisation factor, and no indications for BSM contributions are found.

\begin{figure}
\centering
\begin{overpic}[clip,height=7.5cm]{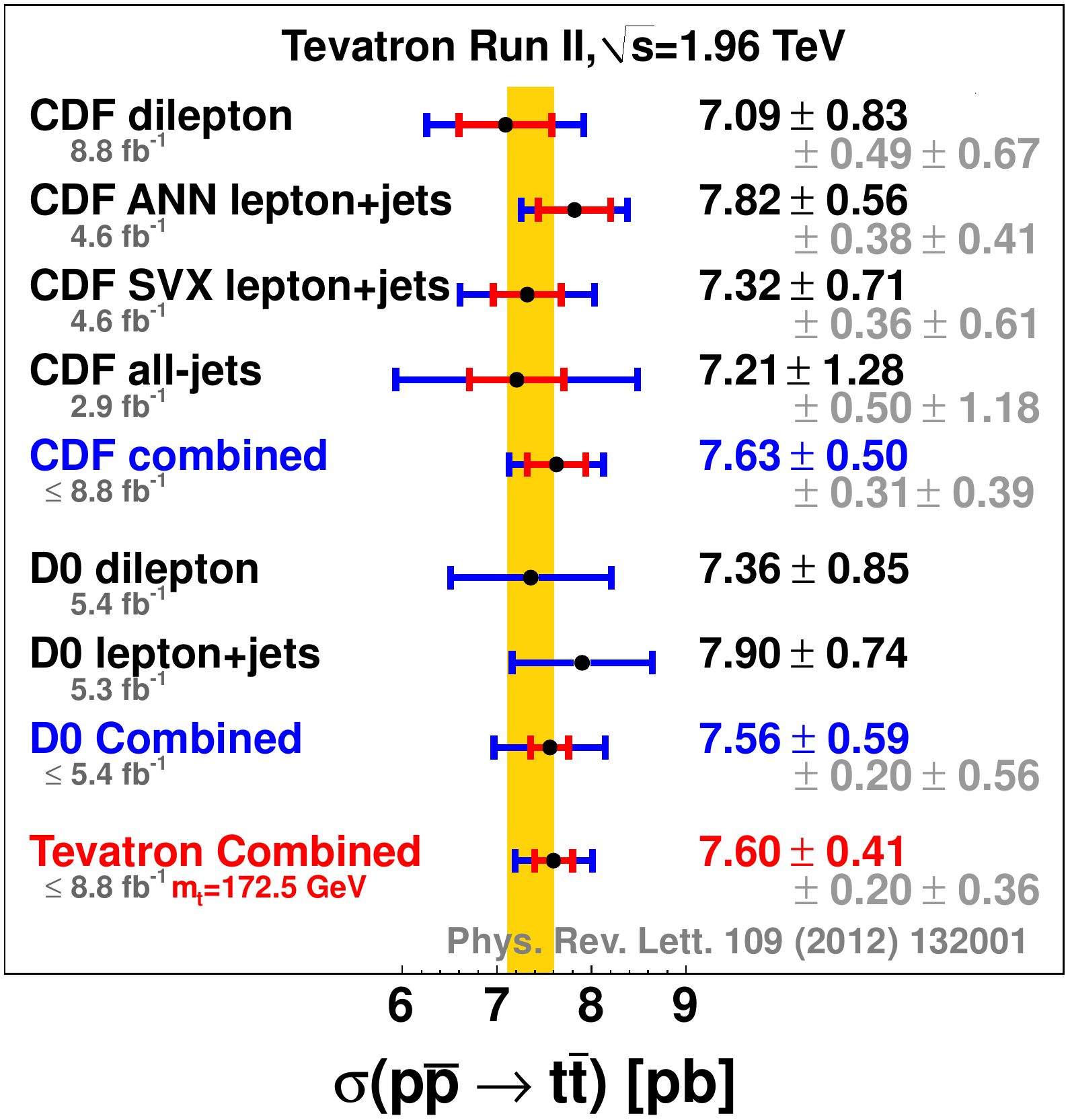}
\end{overpic}
\caption{\label{fig:xsec_tt}
The combination of $\stt$ from CDF and D0 in $p\bar p$ collisions at $\sqrt s=1.96~\TeV$~\cite{bib:xsec_tt}, compared to the SM prediction at NNLO+NNLL~\cite{bib:xsec_tt_nnlo}.
}
\end{figure}

CDF and D0 combined their most precise measurements of \stt in Run II of the Tevatron~\cite{bib:xsec_tt}, summarised in Fig.~\ref{fig:xsec_tt}. The combination is performed using the best linear unbiased estimator (BLUE), taking into account all sources of systematic uncertainties and their correlations between the measurements in the different channels at the two experiments. The combined value of $\stt=7.60\pm0.41$~pb is in good agreement with the SM prediction of $\stt=7.35^{+0.28}_{-0.33}$~pb~\cite{bib:xsec_tt_nnlo}, which is calculated at next-to-next-to-leading order (NNLO) with next-to-next-to-leading logarithmic (NNLL) corrections under the assumption of $\mt=172.5~\GeV$.

\section{Production of single top quarks}

\begin{figure}
\centering
\begin{overpic}[clip,height=4.5cm]{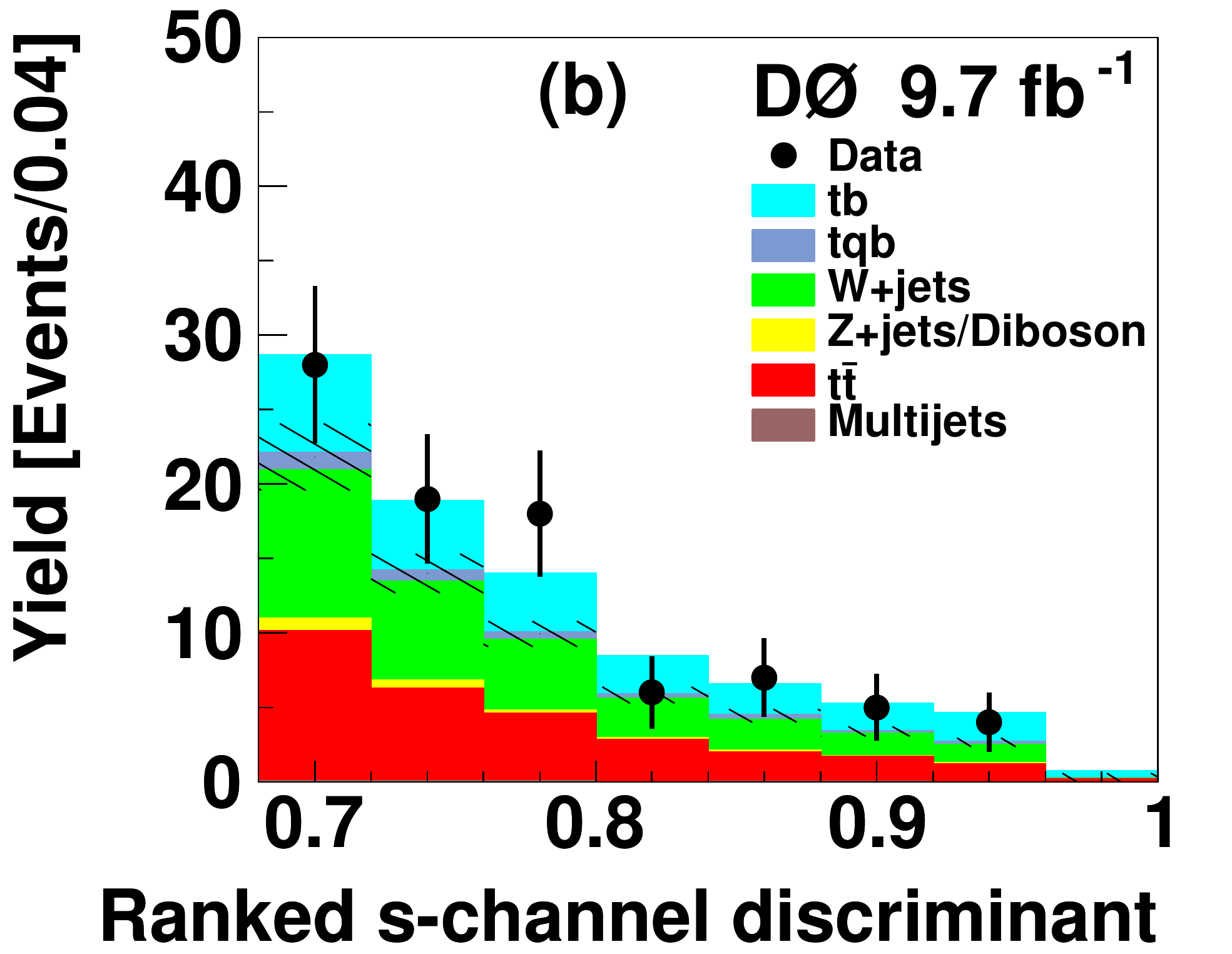}
\put(-1,2){(a)}
\end{overpic}
\qquad
\begin{overpic}[clip,height=4.5cm]{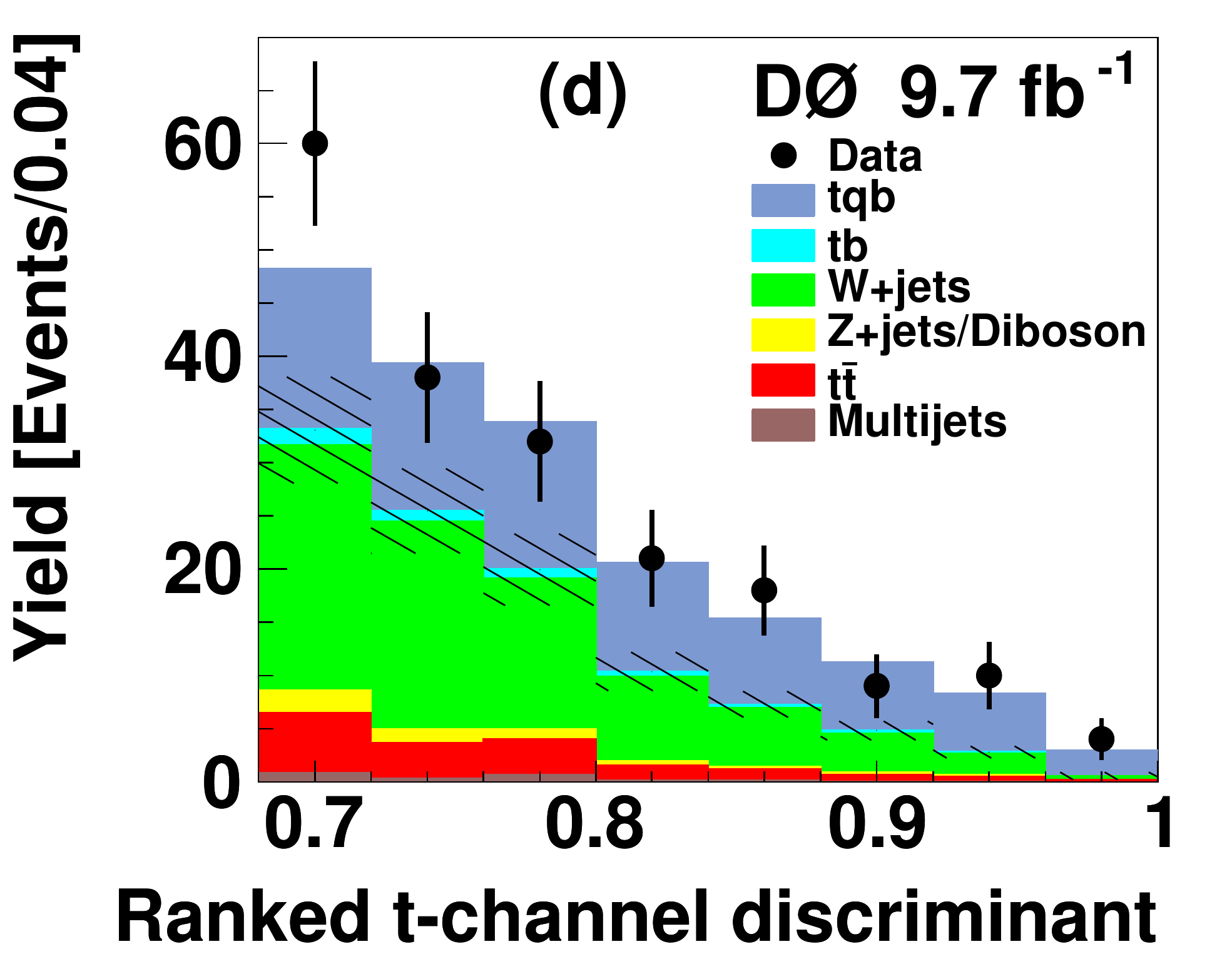}
\put(-1,2){(b)}
\end{overpic}
\caption{\label{fig:st_discr}
{\bf(a)} The ranked $s$-channel discriminant in data compared to simulatinos in the \ljets channel using 9.7~\fb of data~\cite{bib:sch_ev_d0}.
{\bf(b)}~Same as (a), but for the $t$-channel discriminant.
}
\end{figure}

The production of single top quarks in the $s$-channel is difficult to observe at the LHC given its small cross section of $\sigma_{s-{\rm ch.}}=5.5\pm0.2$~pb~\cite{bib:kido}. The experimental situation is less challenging at the Tevatron given that $\sigma_{s-{\rm ch.}}=1.05\pm0.06$~pb contributes about 1/3 of the total $\sigma_t$.

\begin{figure}
\centering
\begin{overpic}[clip,height=5.5cm]{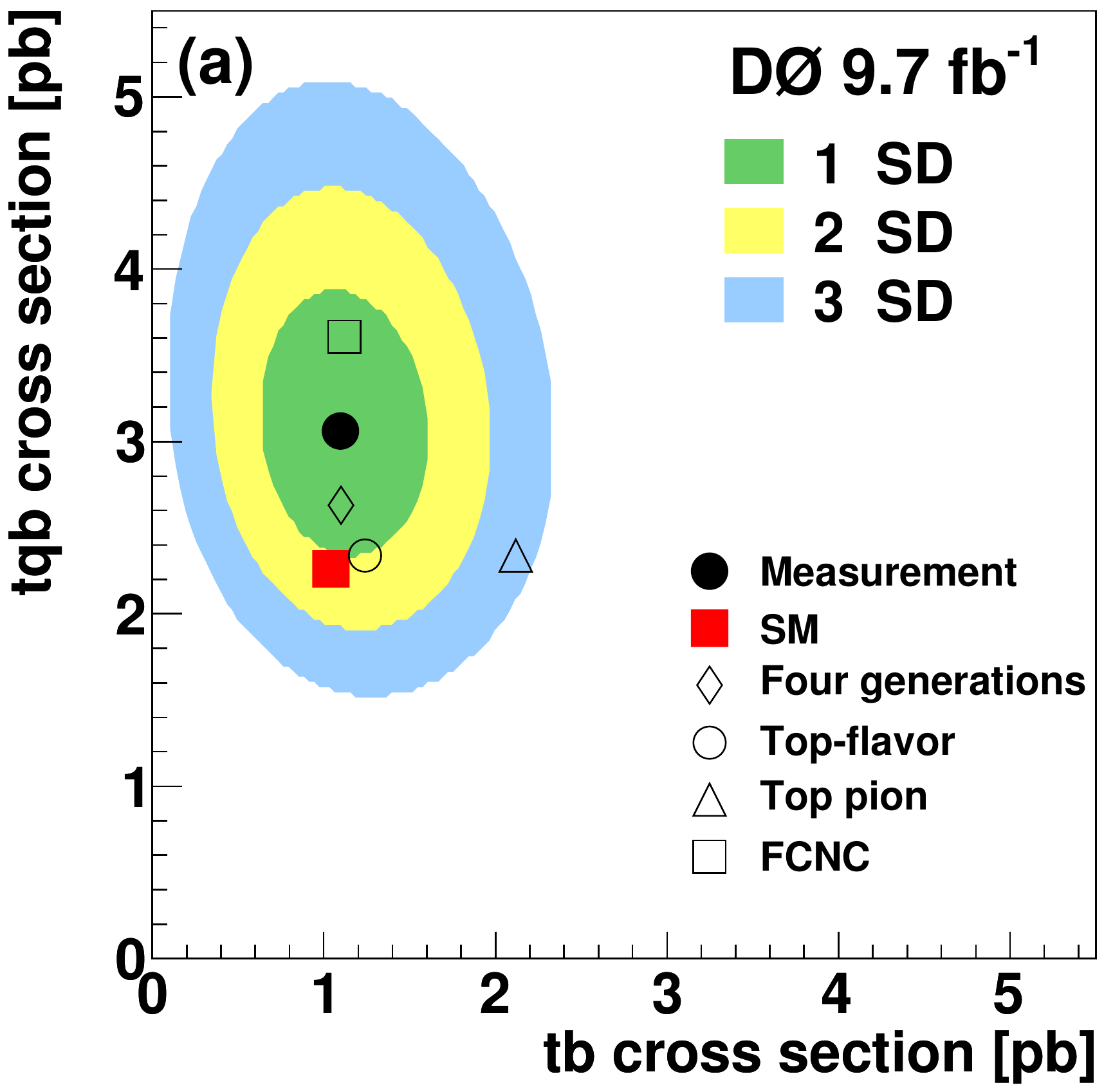}
\put(-1,2){(a)}
\end{overpic}
\qquad
\begin{overpic}[clip,height=5.5cm]{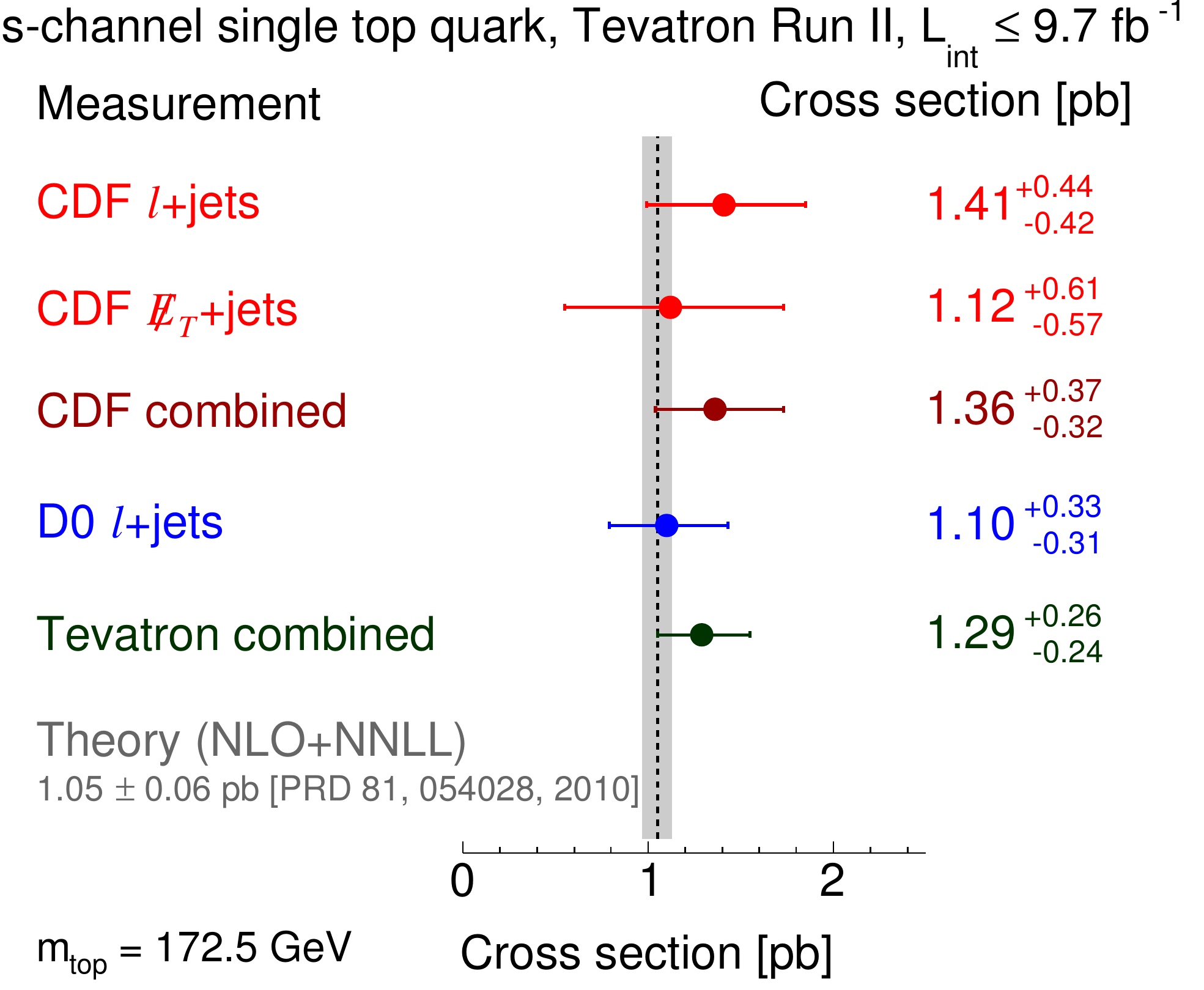}
\put(-1,7){(b)}
\end{overpic}
\caption{\label{fig:st}
{\bf(a)} The simultaneously extracted $\sigma_{s-{\rm ch.}}$ and $\sigma_{t-{\rm ch.}}$ in  the \ljets channel using 9.7~\fb of data~\cite{bib:sch_ev_d0}, compared to the SM expectation and BSM predictions. 
{\bf(b)}~Overview of the measurements of $\sigma_{s-{\rm ch.}}$ at the Tevatron~\cite{bib:sch_obs}.
}
\end{figure}

The first evidence for single top production was achieved by the D0 Collaboration~\cite{bib:sch_ev_d0} in the \ljets channel using 9.7~\fb of data. The sensitivity of the analysis was enhanced by categorising events into signal-enriched and signal-depleted categories according to the jet and $b$-tag multiplicity, where a $b$-tag refers to jets identified as likely to originate from a $b$ quark using a multivariate algorithm. Two discriminants sensitive to the $s$- and $t$-channel, shown in Fig.~\ref{fig:st_discr} were used to extract $\sigma_{s-{\rm ch.}}$ and $\sigma_{t-{\rm ch.}}$ simultaneously. The measured $\sigma_{s-{\rm ch.}}$ and $\sigma_{t-{\rm ch.}}$ are compared to SM expectation and BSM predictions in Fig.~\ref{fig:st}~(a). Without any assumption on $\sigma_{t-{\rm ch.}}$, we measure $\sigma_{s-{\rm ch.}}=1.10\pm0.33$~pb, which agrees with the SM expectation and corresponds to a significance of 3.7 standard deviations (SD). 
CDF and D0 combined their measurements of $\sigma_{s-{\rm ch.}}$ which are shown in Fig.~\ref{fig:st}~(b), and obtained $\sigma_{s-{\rm ch.}}=1.29\pm0.25$~pb~\cite{bib:sch_obs}, in agreement with the SM expectation. This corresponds to an observed (expected) significance of 6.3 (5.1)~SDs, and constitutes a discovery of single top production in the $s$-channel.

\section{Forward-backward asymmetry in $\boldsymbol{t\bar t}$ events}

\begin{figure}
\centering
\begin{overpic}[clip,height=4.5cm]{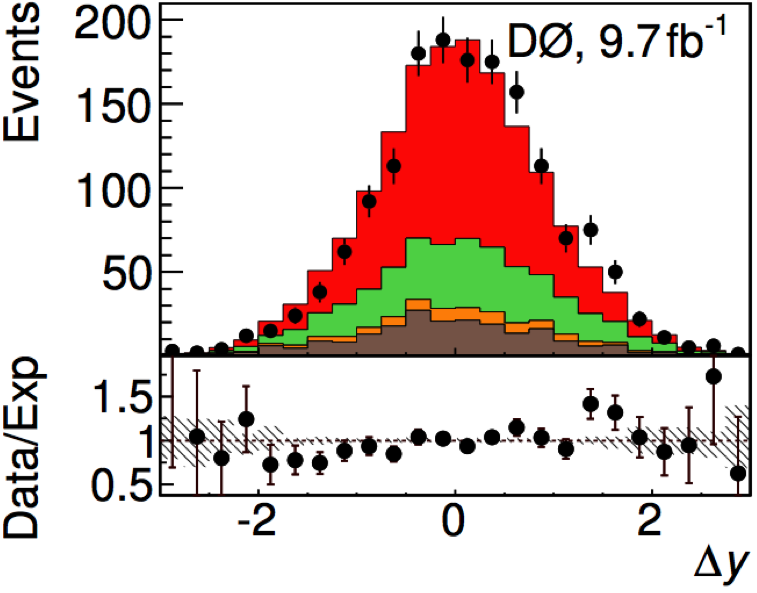}
\put(5,2){(a)}
\end{overpic}
\begin{overpic}[clip,height=4.5cm]{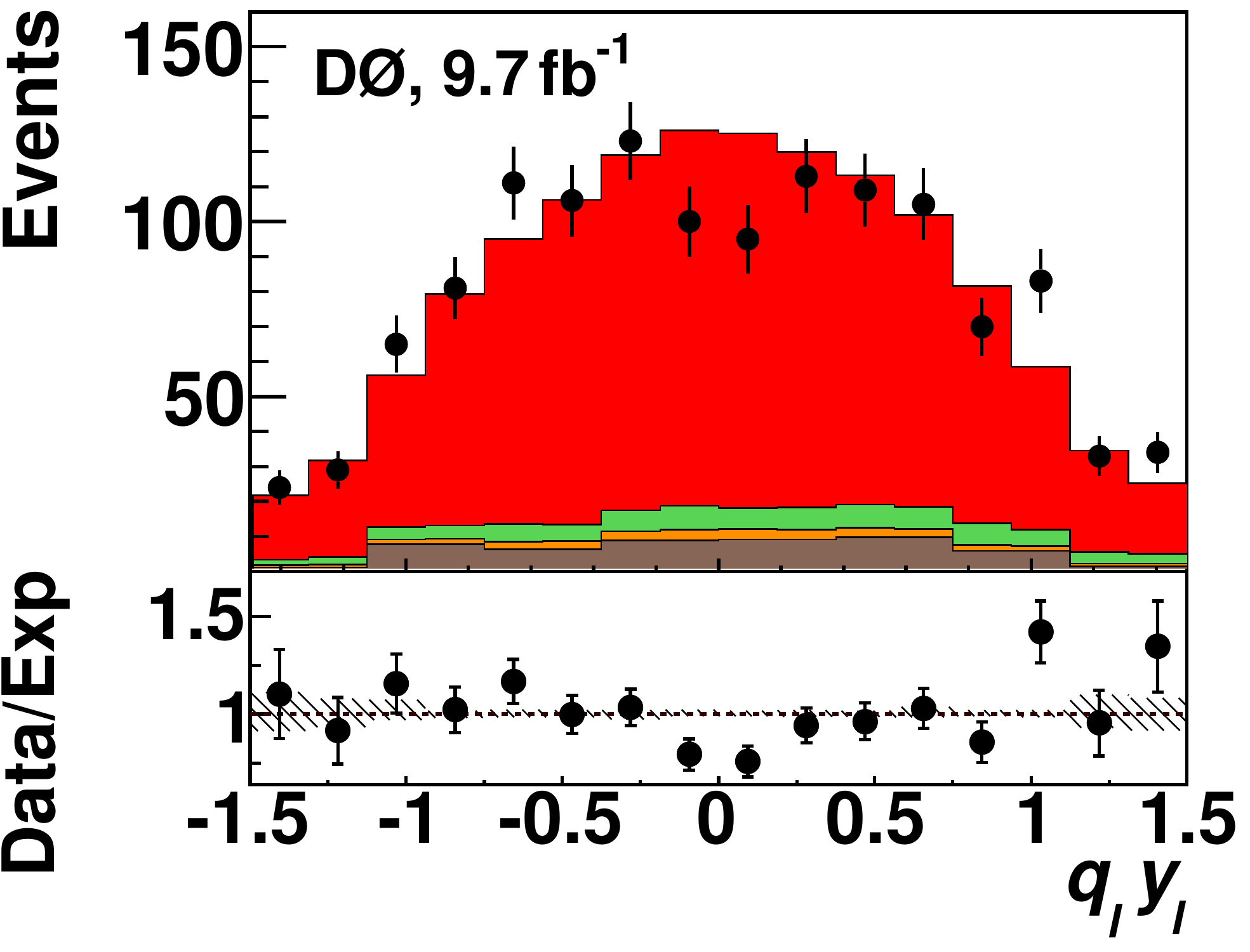}
\put(5,2){(b)}
\end{overpic}
\caption{
\label{fig:afb_meas}
{\bf(a)}~The distribution in $\Delta y$ in data compared to simulations in the \ljets channel using 9.7~\fb of D0 data~\cite{bib:afbtt_d0_lj}.
The data prefers a larger asymmetry than MC@NLO that is used to simulate \ttbar events.
{\bf(b)}~The distribution in $q_\ell y_\ell$ in the laboratory frame in the \ljets channel for the category with four jets and $\geq2$ b-tags~\cite{bib:afbl_d0_lj}.
}
\end{figure}

In the SM, the pair production of top quarks in $p\bar p$ collisions exhibits a forward-backward asymmetry 
\[
\afbtt\equiv\frac{N^{\Delta y>0}-N^{\Delta y<0}}{N^{\Delta y>0}+N^{\Delta y>0}}
\]
of $\approx10\%$ at NNLO with NNLL corrections in the $\ttbar$ rest frame~\cite{bib:afbtt_nnlo}, where $\Delta y\equiv y_t-y_{\bar t}$ and $y_t=\frac12\ln\frac{E_t+p_{z,t}}{E_t-p_{z,t}}$ is the rapidity of the $t$ quark. This asymmetry arises from a negative contribution from the interference of initial and final state radiation, and a larger positive contribution from the interference of box and tree-level diagrams. The Tevatron experiments are in a unique position to measure \afbtt due to the $p\bar p$ initial state.

\begin{figure}
\centering
\begin{overpic}[clip,height=4.5cm]{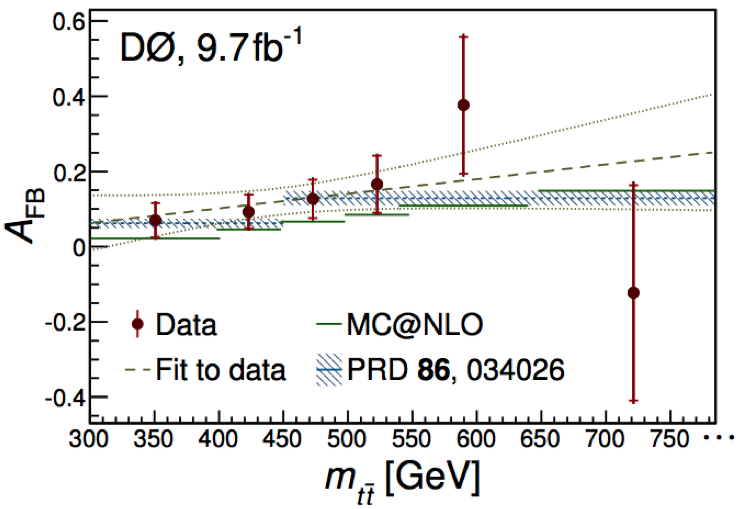}
\put(5,2){(a)}
\end{overpic}
\begin{overpic}[clip,height=4.5cm]{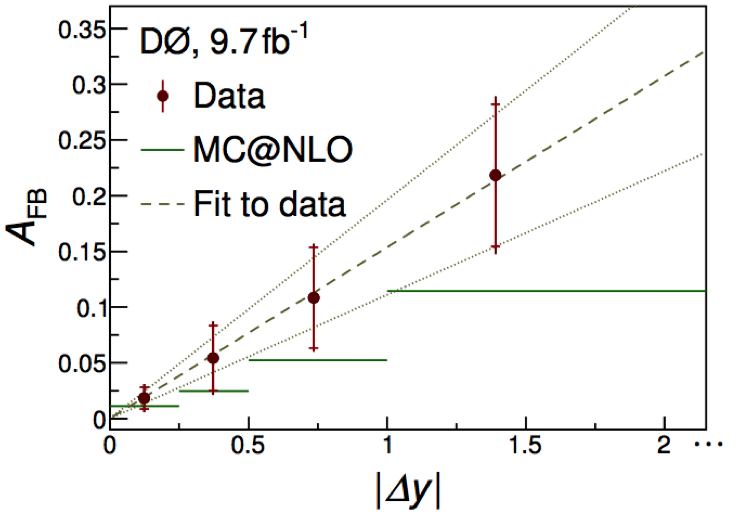}
\put(-1,2){(b)}
\end{overpic}
\caption{
\label{fig:afb_diff}
{\bf(a)} The distribution in $\afbtt$ in the \ttbar\ rest frame versus $M_{\ttbar}$, after background subtraction and corrected for experimental effects from D0 experiment~\cite{bib:afbtt_d0_lj}, is compared to SM expectation calculated at NLO~\cite{bib:afbl_nlo}.
{\bf(b)} Same as~(a), but for the distribution in $|\Delta y|$.
}
\end{figure}

We measured $\afbtt=10.6\pm3.0\%$ in the \ljets channel using events with at least one $b$-tagged jet in 9.7~\fb of data~\cite{bib:afbtt_d0_lj}, as shown in Fig.~\ref{fig:afb_meas}~(a). In this analysis, the kinematic fitter used for the reconstruction of top quarks was extended to cover events with only three jets, which substantially improved the sensitivity of the analysis.
We also investigated the dependence of $\afbtt$ on the invariant mass of the $\ttbar$ system $m_{\ttbar}$ which is compared to SM expectation~\cite{bib:afbl_nlo} in Fig.~\ref{fig:afb_diff}~(a), and on $|\Delta y|$ in~(b). Both measurements agree with predictions within 1.5 SDs.

The lepton-based asymmetry 
\[
\afbl\equiv\frac{N^{q_\ell\eta_\ell>0}-N^{q_\ell\eta_\ell<0}}{N^{q_\ell\eta_\ell>0}+N^{q_\ell\eta_\ell<0}}
\]
is sensitive to $\afbtt$ through the charge-signed pseudorapidity $q_\ell\eta_\ell$ of charged leptons in $t\to\ell\nu b$ decays. The advantage of \afbl over \afbtt is that it does not require any kinematic reconstruction of the top quark and can be easily corrected for detector effects, at the price of a somewhat reduced sensitivity to the forward-backward asymmetry. A representative distribution in $q_\ell\eta_\ell$ in \ljets final states for events with four jets with $\geq2$ $b$-tags is shown in Fig.~\ref{fig:afb_meas}~(b). Our most precise measurement of \afbl is performed in the \ljets channel using 9.7~\fb of data and reads $\afbl=5.0\pm3.6\%$, which is consistent with the SM expectation at NLO~\cite{bib:afbl_nlo}.

\begin{figure}
\centering
\begin{overpic}[clip,height=7cm]{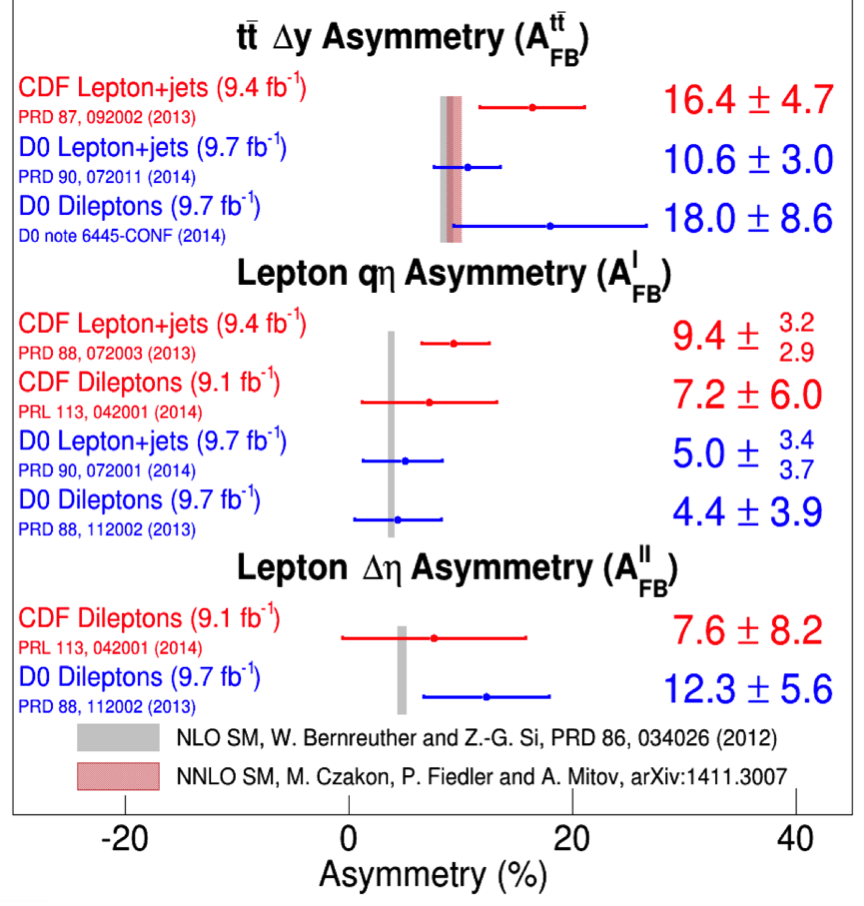}
\end{overpic}
\caption{\label{fig:asymm}
The overview of the measurements of \afbtt, \afbl, and $\afb^{\ell\ell}$~(not discussed here) by CDF and D0, compared to SM expectation. 
The references to the measurements and the theory predictions are given in the plot.
}
\end{figure}

An overview of all measurements of the forward-backward asymmetry from the Tevatron is given in Fig.~\ref{fig:asymm}.

\section{Top quark mass}
The  most precise determination of \mt\ in dilepton final states at the Tevatron is performed by D0 using 5.4~\fb\ of data~\cite{bib:mtopll_d0}. It is a combination of two measurements, using the matrix element (ME) technique~\cite{bib:mtopllme_d0}, which will be described below in the context of the \ljets\ channel, and the neutrino weighting technique~\cite{bib:mtopll_d0}. To reduce the systematic uncertainty, we apply the {\em in situ} jet energy scale (JES) calibration in \ljets\ final states derived in Ref.~\cite{bib:mtoplj_d0}, accounting for differences in jet multiplicity, luminosity, and detector ageing. A combination of both analyses in the dilepton final states yields $\mt=173.9~ \pm 1.9~({\rm stat}) \pm 1.6~({\rm syst})~\GeV$.

\begin{figure}
\begin{centering}
\includegraphics[height=4.5cm]{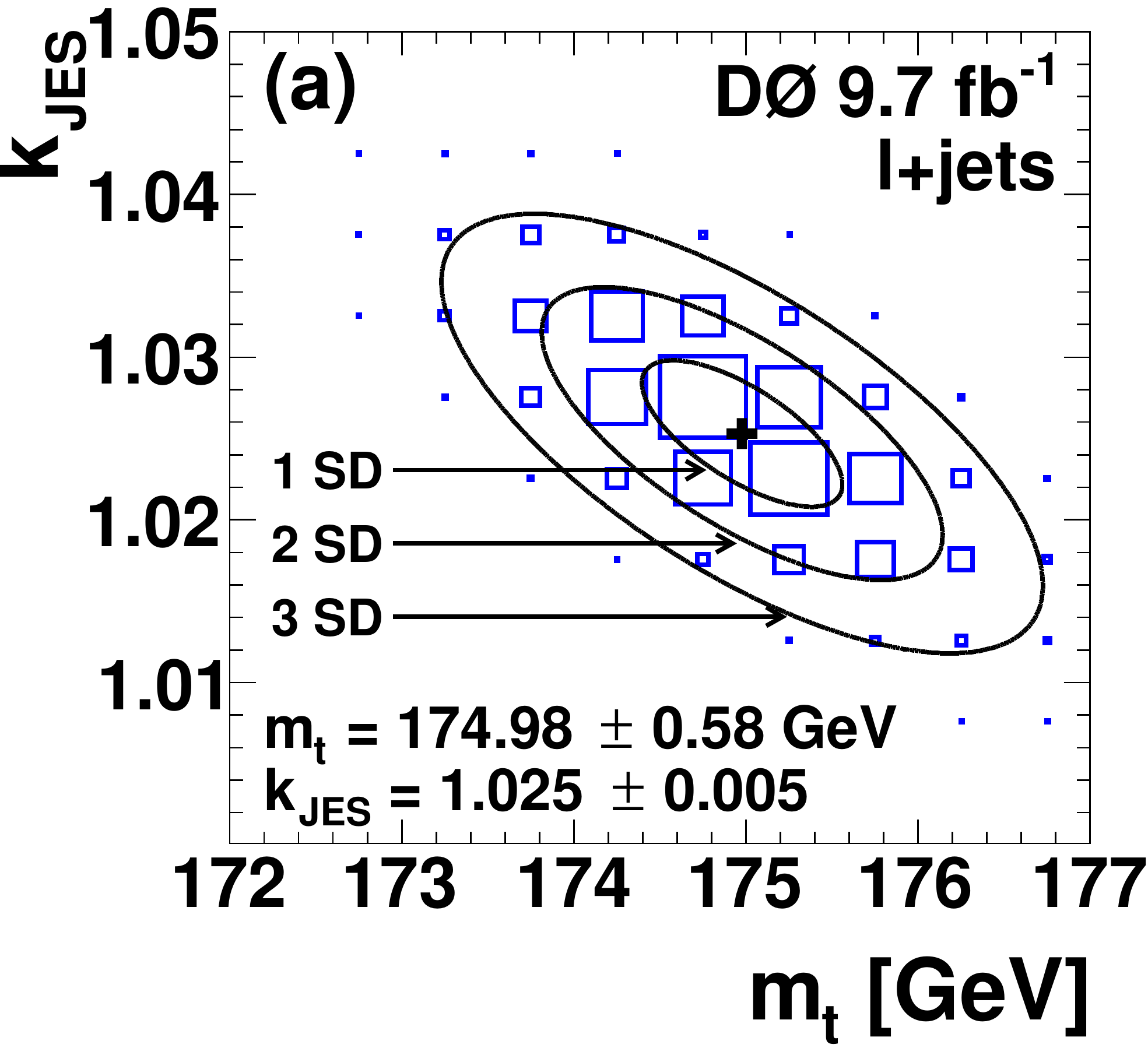}
\includegraphics[height=4.5cm]{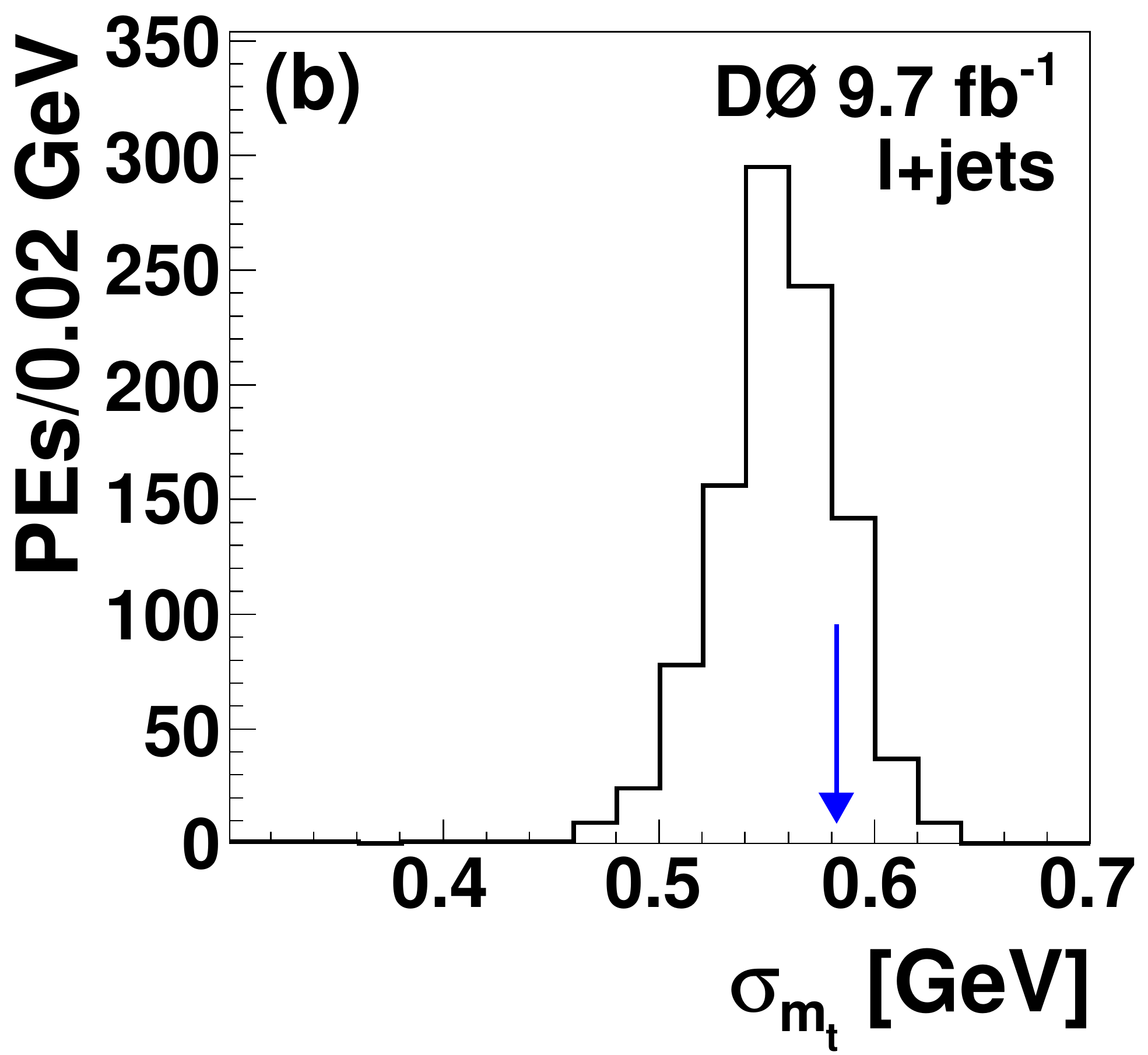}
\par\end{centering}
\caption{
\label{fig:like2d}
{\bf (a)}~Two-dimensional likelihood ${\cal L}(\mt,\kjes)/{\cal L}_{\rm max}$ for data. Fitted contours of equal probability are overlaid as solid lines. The maximum is marked with a cross. Note that the bin boundaries do not necessarily correspond to the grid points on which $\cal L$ is calculated.
{\bf (b)}~Expected uncertainty distributions for \mt with the measured uncertainty indicated by the arrow.
}
\end{figure}

We also perform a measurement of \mt in the \ljets channel using 9.7~\fb of data~\cite{bib:mt} with the ME technique, which determines the probability of observing each event under both the $\ttbar$ signal and background hypotheses described by the respective MEs.
The overall JES is calibrated {\it in situ} by constraining the reconstructed invariant mass of the hadronically decaying $W$ boson to $\mw=80.4$~GeV. 
 Jet energies are corrected to the particle level using calibrations derived from exclusive $\gamma+$jet, $Z+$jet, and dijet events~\cite{bib:jes}. These calibrations account for differences in detector response to jets originating from a gluon, a $b$~quark, and $u,d,s,$ or $c$~quarks. The analysis was performed blinded in \mt. 
%
Applying the ME technique to data, the result considering only statistical uncertainties and the two-dimensional likelihood distribution in $(\mt,\kjes)$ is shown in Fig.~\ref{fig:like2d}~(a), where the statistical uncertainty on \mt of 0.58~\GeV also includes the statistical contribution from the overall JES factor \kjes. Figure~\ref{fig:like2d}~(b) compares the measured total statistical uncertainty on \mt with the distribution of this quantity from the pseudoexperiments at $\mt^{\rm gen}=172.5~\GeV$ and $\kjes^{\rm gen}=1$. Systematic uncertainties are evaluated for three categories:
(i)~modelling of signal and background events,
(ii)~uncertainties in the simulation of the detector response,
(iii)~and uncertainties associated with procedures used and assumptions made in the analysis. The largest contribution comes from category~(i), followed by category~(ii). The dominant uncertainties are from the modelling of hadronisation and underlying event~(0.26~\GeV), the modelling of the hard  ME~(0.15~\GeV), the residual dependence of JES on $\eta$ and $\pt$ of the jet~(0.21~\GeV), and the flavour-dependence of the calorimeter response to jets~(0.16~\GeV).
In summary, we measure $\mt = 174.98 \pm 0.58\thinspace({\rm stat+JES}) \pm 0.49\thinspace({\rm syst})~\GeV$ with the ME technique in \ljets final states, which is consistent with the values given by the current world combination of \mt~\cite{bib:combiworld} and achieves by itself a similar precision. With an uncertainty of $0.43\%$, it is the most precise single measurement of \mt.

\begin{figure}
\begin{centering}
\includegraphics[height=8cm]{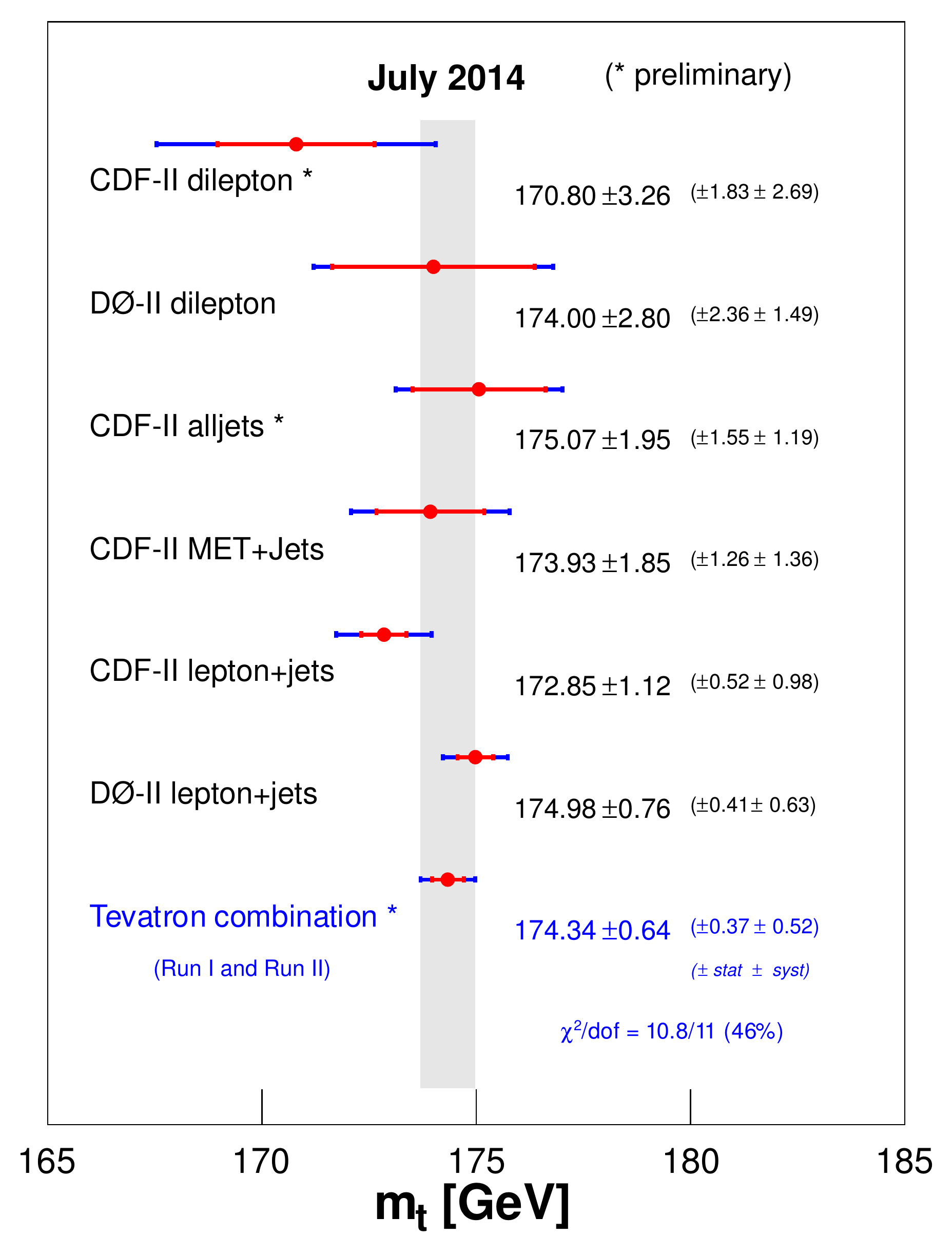}
\par\end{centering}
\caption{
\label{fig:combi}
Summary of the measurements performed in Run~II at the Tevatron which are used as inputs to the Tevatron combination~\cite{bib:mt_tev}. 
The Tevatron average value of \mt obtained using input measurements from Run~I and Run~II is given at the bottom, and its uncertainty is shown as a gray band. 
}
\end{figure}

Our results are included in the Tevatron combination from July 2014~\cite{bib:mt_tev}, which is performed taking into account 10 published and 2 preliminary results from the CDF and D0 collaborations using $p\bar p$ collision data from Run~I and Run~II of the Fermilab Tevatron Collider. Taking into account correlations between considered sources of systematic uncertainty, the final result reads $\mt=174.34\pm0.64~\GeV$ corresponding to a precision of 0.37\%, with a relative contribution from our  measurement in \ljets final states of 67\%. An overview of input measurements performed using Run~II data is presented in Fig.~\ref{fig:combi}. The consistency of the input measurements is given by the value of the $\chi^2$ distribution for 11 degrees of freedom and corresponds to a $\chi^2$ probability of 46\%.

\section{Conclusions}
I presented recent measurements of the production of \ttbar pairs and single top quarks by the D0 experiment and the respective Tevatron combinations, which are found in agreement with the SM expectations. The forward-backward asymmetry in the \ttbar system displays some tension between the theory and experiment in differential distributions at the level of 1.5 standard deviations. The world's most precise single measurement of the top quark mass has been presented.

\section*{Acknowledgments}
I would like to thank my colleagues from the D0 experiment for their help in preparing this article. I also thank the staffs at Fermilab and collaborating institutions, as well as the D0 funding agencies.


\begin{thebibliography}{99}
\bibitem{bib:foot1}
Here and in the following, the contribution of the charge-conjugated process and higher order corrections is not written explicitly.
\end{thebibliography}
\end{document}